\documentclass{amsart}
\usepackage{graphicx}
\usepackage{color}
\def\br{{\bf r}}
\UseRawInputEncoding
\def\dfrac#1#2{\displaystyle{#1\over #2}}

\def\bv{{\bf V}}
\def\bV{{\bf V}}

\def\Div{\mbox{div}\,}
\def\Rot{\mbox{curl}\,}

\def\bB{{\bf B}}
\def\bx{{\bf x}}
\def\bE{{\bf E}}

\title[Cold plasma in a constant magnetic field]{On plane oscillations of the cold plasma in a constant magnetic field} % Paper title
\subjclass{Primary 35Q60; Secondary 35L60, 35L67, 34M10} % UDC number
\author{Olga S. Rozanova} % Author's first and last names
\address{Mathematics and Mechanics Department, Lomonosov Moscow State University, Leninskie Gory,
Moscow, 119991,
Russian Federation} % Author's address
\email{rozanova@mech.math.msu.su} % Author's e-mail
\theoremstyle{plain}
\newtheorem{theorem}{Theorem}
\newtheorem{lemma}{Lemma}
\theoremstyle{definition}
\newtheorem{remark}{Remark}
\numberwithin{equation}{section}
\numberwithin{theorem}{section}
\numberwithin{lemma}{section}
\numberwithin{definition}{section}
\numberwithin{remark}{section}

\begin{document}
\begin{abstract}
We consider a class of two-dimensional solutions of the cold plasma equations compatible with a constant magnetic field and a constant electric field. For this class, under various assumptions about the electric field, we study the conditions on the initial data that guarantee the global existence of the classical solution of the Cauchy problem for a given period of time or a finite blowup. Particular attention is paid to the class of solutions with axial symmetry.
\end{abstract}
\maketitle
%\tableofcontents

% Text of the paper

\section{Introduction}

%\bigskip
Plasma is actually a two-phase medium consisting of ions and electrons interacting with each other. There are many models that describe its behavior under various modes (see, e.g.,~\cite{ABR78},  \cite {GR75}). Among them, the model of the so-called cold (or electron) plasma, which includes the movement of only electrons, stands out. It is believed that plasma at low temperatures obeys such a model, which justifies the term "cold plasma". At present, cold plasmas are being intensively studied in connection with electron accelerators in the wake wave of a powerful laser pulse \cite{esarey09}.

The equations of hydrodynamics of a cold plasma in the
non-relativistic approximation in dimensionless quantities take the
form
\begin{eqnarray}\label{base1.1}
\dfrac{\partial n }{\partial t} + \Div(n \bv)=0,\quad
\dfrac{\partial \bv }{\partial t} + \left( \bv \cdot \nabla \right) \bv
= \, - \bE -  \left[\bv \times  \bB\right],\\
\frac{\partial \bE }{\partial t} =   n \bv
 + {\rm rot}\, \bB,\qquad %\label{base1.3}\\
\frac{\partial \bB }{\partial t}  =
 - {\rm rot}\, \bE,\qquad \Div \bB=0,\label{base1.5}
\end{eqnarray}
$ n$ and $ \bv=(V_1, V_2, V_3)$  are the density and velocity of electrons,
$ \bE=(E_1, E_2, E_3)$ and  $  \bB=(B_1, B_2, B_3) $  are vectors of electric and magnetic fields.
All components of solution depends on $t\in {\mathbb R}_+$ and $x\in{\mathbb R}^3$.
The ions in this model are assumed to be immobile.

The main problem that physicists are interested in in connection with the equations describing cold plasma is to determine the conditions on the initial data under which the solution retains the original smoothness for as long as possible (ideally, always). It is believed that during the formation of a singularity of a smooth solution, energy is released that heats the plasma, so that the assumption of immobility of the ions ceases to be valid.

For the model case of one spatial variable, which is nevertheless very important for testing numerical methods \cite{CH18}, the original system of equations is greatly simplified. The problem of the formation of singularities in this case is currently quite well studied (\cite{RChZAMP21}), including special reductions that make it possible to trace the influence of the magnetic field in the so-called Davidson model (\cite{david72}, \cite{RChZAMP22}).

However
 system \eqref{base1.1}, \eqref{base1.5}  in the space of many spatial variables is extremely complex and includes many modes of oscillations. In particular, the two-dimensional case is important from the point of view of physical experiments. As for numerical studies, there are results confirming the complex behavior of the medium \cite{GFCA}.

 Up to now, for the case of many spatial dimensions there exit theoretical results only for the case of electrostatic (i.e. ${\rm rot}\, \bE =0$) oscillations  \cite{Roz22_MMAS}, for the solution with the radial symmetry \cite{R_PhysicaD22} or linear dependence on the space variables (the affine solutions) \cite{RTur}. For the case $\bB=0$.

In this paper, we study a particular case of two-dimensional (plane) oscillations for which the magnetic field is a nonzero constant. In other words,
 $ \bv=(V_1, V_2, 0)$,
$ \bE=(E_1, E_2, 0)$,  $  \bB_0=(0,0, B_0) $, and  $V_1, V_2, E_1, E_2, n$ depend on $x_1, x_2, t$. If the magnetic is constant, then
\begin{equation*}\label{cond}
\Rot \bE=0,\quad  \Rot (n \bv)=0.
\end{equation*}
As follows from the second equation of \eqref{base1.1},
%For the case of plain oscillations $\bv=\bv(x_1,x_2)$, $\bE=\bE(x_1,x_2)$, $V_3=E_3=0$, $\xi_1=\xi_2=0$, and we have
\begin{eqnarray}\label{xi}
\dfrac{\partial \xi }{\partial t}  &+& ( \bv \cdot \nabla)  \xi  = - {\mathcal D}(\xi+B_0),
\end{eqnarray}
where
${\mathcal D}=\Div \bv$, $\Rot \bv=(0, 0,
\xi)$. Thus,
for the case  $B_0\ne 0$ the condition  $\Rot (n \bv)=0$ generally does not hold for all $t\ge 0$. To avoid this problem, we suppose
 $n=0$. Then the first equations in \eqref{base1.1} and \eqref{base1.5} are satisfied identically for any stationary $\bE=\bE_0(x_1,x_2)$,
 such that $ {\rm rot}\, \bE_0=0$.

Of course, one can argue about whether the considered class of solutions of the cold plasma equations has a physical meaning. However, from a mathematical point of view, the study of motion in a given landscape of electric and magnetic fields is extremely interesting. In a sense, this problem resembles the problem of fluid motion on a rotating plane, which arises in geophysical applications \cite{RU}, but is much more complicated. In particular, as will be shown below, an increase in the magnetic field generally leads to a smoothing of the solution.

Thus, the system under consideration is
\begin{eqnarray}\label{1}
\dfrac{\partial \bv }{\partial t} + \left( \bv \cdot \nabla \right) \bv
= \, - \bE_0 -  \left[\bv \times  \bB_0\right],
\end{eqnarray}
together with
 the initial data
\begin{equation}\label{CD1}
\bv |_{t=0}=
\bv_0, (x_1, x_2)  \in C^2({\mathbb R}^2). %\mbox{such\, that}
\end{equation}
For the sake of simplicity we assume $B_0>0$.

The vectorial equation  \eqref{1} has the following differential implications.

1. Matrix equation for unknown matrix of derivatives $Q$:
\begin{eqnarray*}\label{2}
\dfrac{\partial {\mathcal V} }{\partial t} + \left( \bv \cdot \nabla \right) {\mathcal V}
= \, -{\mathcal V}^2-  B_0 L {\mathcal V} - S_0(x_1,x_2),
\end{eqnarray*}
where
\begin{eqnarray}\label{Q}
{\mathcal V}=(v_{ij})=\left(\partial_{x_i}{V_j}\right),\quad S_0=(s_{ij})=\left(\partial_{x_i}{E_{0j}}\right), \quad \partial_{x_i}{E_{0j}}= \partial_{x_j}{E_{0i}},\quad i,j=1,2,\quad  L=
\begin{pmatrix}
  0 & 1\\ -1 & 0
\end{pmatrix}.
\end{eqnarray}

2.A pair of scalar equations for $\mathcal D$ and $\xi$:
\begin{eqnarray}\label{D}
\dfrac{\partial {\mathcal D}}{\partial t} &+& ( \bv \cdot \nabla
{\mathcal D} ) =   - {\mathcal D}^2 + 2J -\lambda(x_1, x_2) - B_0 \xi,
\end{eqnarray}
where $J=\det(\|\partial_{x_i}{V_j}\|),$ $i,j=1,2$, $\lambda=\Div \bE$, and \eqref{xi}.

We see that the equations \eqref{1}, \eqref{Q}, \eqref{D}, \eqref{xi} are written along the same characteristic field
\begin{eqnarray}\label{x}
\dfrac{\partial x_i}{\partial t} &+& ( \bv \cdot \nabla)
x_i =  V_i, \quad i=1,2,\quad (x_1(0), x_2(0))=(x_{01}, x_{02}),
\end{eqnarray}
therefore for $\dfrac{d}{dt}=\dfrac{\partial }{\partial t} + \bv \cdot \nabla$ the hyperbolic system \eqref{1}, \eqref{Q}, \eqref{x} can be considered as a closed quadratically nonlinear ODE system  for the vectors $\bv$, $\bx=(x_1, x_2)$, and matrix ${\mathcal V}$. Formation of a singularity means a finite time blow-up of a component of $Q$ for least one initial point $(x_{01}, x_{02})$.

Obviously, for an arbitrary $\bE_0$, the system of 8 equations
\begin{eqnarray}%\label{ODE}
\dfrac{ d \bv }{d t}
&=& -  B_0 L \bv - E_0(x_1,x_2),\label{ODEv}\\\nonumber
\dfrac{d \bx}{d t} &=&  \bv, \\%\label{ODEx}
\dfrac{ d{\mathcal V} }{d t}
&=& \, -{\mathcal V}^2-  B_0 L {\mathcal V} - S_0(x_1,x_2),\label{ODEq}\\
 ((V_1(0), V_2(0), x_1(0), x_2(0), Q(0))&=&(V_1(x_{01}, x_{02}), V_2(x_{01}, x_{02}), x_{01}, x_{02},\left(\partial_{x_i}{V_j(x_{01}, x_{02})}\right)),\nonumber
\end{eqnarray}
$i,j=1,2,$
can be solved only numerically.

Nevertheless, for a specific choice of $\bE_0=\mathcal S_0 \bx$ with a constant symmetric matrix $\mathcal S_0=(s_{ij})$, $i,j=1,2$, one can obtain a criterion for the formation of singularities and a sufficient condition for the global in $t$ smoothness of solution in the terms of the initial data  $\bv_0$ and input parameters $s_{ij}$ and $B_0$, see Sec.\ref{S2}, Theorem \ref{T1}.

For the general case, the sufficient conditions for the smoothness look cumbersome, so we present their consequences for the case of axial symmetry
 \begin{eqnarray}\label{axial}
\bv = U(r) \bx +V(r) \bx_\bot, \quad \bE_0= S(r) \br,\quad r=\sqrt{x_1^2+x_2^2}, \quad \bx_\bot=(x_2,-x_2),
\end{eqnarray}
see Sec.\ref{S2.1}.

In section \ref{S3} we study the axisymmetric case with variable $\bE_0$ such that $S_- \le S(r)\le S_+ $ and $\lambda_- \le {\rm div } \bE_0 ( r)\le \lambda_+ $, with constants $S\pm$ and $\lambda\pm$
   and find sufficient conditions for
   $\bv_0$, which guarantees the classical smoothness of the Cauchy problem on a period depending on $B_0$ and $S(r)$, see Theorem \ref{T3}.

\medskip

\section{The case of affine $\bE_0$.}\label{S2}

\medskip

It is easy to see that in this case the matrices $S(x_1,x_2)=\mathcal S_0$ do not depend on $(x_1, x_2)$, so the \eqref{ODEq} system can be considered separately.

Let us show that \eqref{ODEq} can be linearized.
We need the following version of the Radon lemma (1927)
\cite{Riccati}, Theorem 3.1, see also \cite{Radon}.

\begin{theorem}[The Radon lemma]
\label{T2} A matrix Riccati equation
\begin{equation}
\label{Ric}
 \dot W =M_{21}(t) +M_{22}(t)  W - W M_{11}(t) - W M_{12}(t) W,
\end{equation}
 {\rm (}$W=W(t)$ is a matrix $(n\times m)$, $M_{21}$ is a matrix $(n\times m)$, $M_{22}$ is a matrix  $(m\times m)$, $M_{11}$ is a matrix  $(n\times n)$, $M_{12} $ is a matrix $(m\times n)${\rm )} is equivalent to the homogeneous linear matrix equation
\begin{equation}
\label{Lin}
 \dot Y =M(t) Y, \quad M=\left(\begin{array}{cc}M_{11}
 & M_{12}\\ M_{21}
 & M_{22}
  \end{array}\right),
\end{equation}
 {\rm (}$Y=Y(t)$  is a matrix $(n\times (n+m))$, $M$ is a matrix $((n+m)\times (n+m))$ {\rm )} in the following sense.

Let on some interval ${\mathcal J} \in \mathbb R$ the
matrix-function $\,Y(t)=\left(\begin{array}{c}Q(t)\\ P(t)
  \end{array}\right)$ {\rm (}$Q$  is a matrix $(n\times n)$, $P$  is a matrix $(n\times m)${\rm ) } be a solution of \eqref{Lin}
  with the initial data
  \begin{equation*}\label{LinID}
  Y(0)=\left(\begin{array}{c}I\\ W_0
  \end{array}\right)
  \end{equation*}
   {\rm (}$ I $ is the identity matrix $(n\times n)$, $W_0$ is a constant matrix $(n\times m)${\rm ) } and  $\det Q\ne 0$ on ${\mathcal J}$.
  Then
{\bf $ W(t)=P(t) Q^{-1}(t)$} is the solution of \eqref{Ric} with
$W(0)=W_0$ on ${\mathcal J}$.
\end{theorem}

System \eqref{Q} can be written as \eqref{Ric} with
%$$
\begin{eqnarray*}\label{M}
W={\mathcal V}, \quad
M_{11}=\begin{pmatrix}
   0 & 0\\ 0 & 0
\end{pmatrix},\quad
 M_{12}=\begin{pmatrix}
  1 & 0\\ 0 & 1
\end{pmatrix},\quad
M_{21}=-\mathcal S_0,\quad
M_{22}=-B_0 L.\\\nonumber
%$$$$
\end{eqnarray*}

Thus, we obtain the linear Cauchy problem
\begin{eqnarray}\label{ode}
\label{matr}
 \begin{pmatrix}
  \dot q_{11}&\dot q_{12}\\
 \dot q_{21}&\dot q_{22}\\
  \dot p_{11}&\dot p_{12}\\
 \dot p_{21}&\dot p_{22}\\
\end{pmatrix}
=M
\begin{pmatrix}
   q_{11}& q_{12}\\
  q_{21}& q_{22}\\
   p_{11}& p_{12}\\
  p_{21}& p_{22}\\
\end{pmatrix},\quad
M=\begin{pmatrix}
0& 0&1&0\\
0& 0&0&1\\
-s_{11}&-s_{12}&0& -B_0\\
-s_{21}&-s_{22}&B_0& 0\\
\end{pmatrix},
\end{eqnarray}
subject to initial conditions
\begin{eqnarray}\label{CCP}
\begin{pmatrix}
   q_{11}& q_{12}\\
  q_{21}& q_{22}\\
   p_{11}& p_{12}\\
  p_{21}& p_{22}\\
\end{pmatrix}(0)=\begin{pmatrix}
  1& 0\\
  0& 1\\
   v_{11}& v_{12}\\
  v_{21}& v_{22}\\
\end{pmatrix}.
\end{eqnarray}
It is a linear system with constant coefficients that can be solved explicitly. Recall that ${\rm det} Q(0)=1$. Thus, the derivatives $v_{ij}$, $i,j=1,2$, remain bounded for all $t>0$ if and only if ${\rm det} Q>0$ for all $t >$0. If ${\rm det} Q>0$ for all $t>0$ for any characteristic starting from $(x_{01}, x_{02}) \in \mathbb R^2 $, then the solution of the Cauchy problem is the problem preserves smoothness for all $t>0$.

Nevertheless, this criterion is implicit, and it would be more convenient to find a sufficient condition guaranteeing global smoothness, i.e., investigate when ${\rm det} Q>0$ for all $t>0$.

The eigenvalues of $M$ are as follows:
\begin{eqnarray*}
\mu_{1234}&=& \pm \frac{1}{\sqrt{2}}\, \sqrt{\pm \sqrt{(B_0^2+\lambda)^2-4 K}-({B^2+\lambda})}, \\
\lambda &=& \Div \bE_0 = s_{11}+s_{22}, \quad K=\det(\partial_{x_i}{E_{0j}})=s_{11} s_{22}- s^2_{12}.
\end{eqnarray*}
First of all, note that if $\Re \,\mu_{i}\ne 0$, $i=1,\dots, 4$, then there is no choice $P=(v_{ij} (0),s_{ ij}(0))\in \mathbb R^8 $, which guarantees the positivity of ${\rm det} Q$ such that this positivity also holds in a neighborhood of $P$. Indeed, for the case $\Re \,\mu_{i}\ne 0$ the solution $q_{ij}(t)$, generally speaking, contains an increasing exponent.

Therefore, to find a sufficient smoothness condition that is stable in the initial data, we focus on the case $\Re \,\mu_{i}= 0$. It is easy to verify that it is satisfied if and only if
\begin{eqnarray}\label{mu}
  4K&<& (B_0^2+\lambda)^2, \quad K>0.
\end{eqnarray}

The next condition, necessary for the  boundedness of ${\rm det} Q$, is that the frequencies $|\mu_i|$ are not resonant, i.e.
\begin{eqnarray}\label{nr}
 \frac{\omega_-}{\omega_+}&\ne&\frac{m}{n},\qquad n\in\mathbb N, \, m\in \mathbb Z,\\
\omega_\pm &=& \frac{1}{\sqrt{2}}\,{\sqrt{ {B^2+\lambda}\pm\sqrt{(B_0^2+\lambda)^2-4 K}}}.\nonumber
\end{eqnarray}

It can be explicitly calculated that
\begin{eqnarray}\label{detQ}
&&{\rm det} Q = \\&&\frac{1}{k} \left[C + A_- \sin (\omega_+-\omega_-) t + B_- \cos (\omega_+-\omega_-) t + A_+ \sin (\omega_++\omega_-) t + B_+ \cos (\omega_+-\omega_-) t \right],\nonumber
\end{eqnarray}
with constant $k, C, A_\pm, B_\pm$ depending on $v_{ij}(0),s_{ij}(0), B_0$ (in a rather cumbersome way). It is obvious that
$C+A_-+A_+=k$.

Here
\begin{eqnarray*}\label{C}
C= {B_0 K} \sqrt{(B_0^2+\lambda)^2-4 K} \left[B_0^3 +(v_{12}-v_{21}) B_0^2 + (\lambda + 2 J(0)) B_0 + 2 s_{12} (v_{11}- v_{22})- (v_{12}-v_{21}) (s_{11}-s_{22})\right],
\end{eqnarray*}
\begin{eqnarray*}\label{AB}
A_-&=& \lambda (\lambda + 2 B^2) \left[ a_-(\omega_-+\omega_+) + b_-(\omega_--\omega_+)\right],\quad B_-= \lambda (\lambda + 2 B^2) \left[ b_-(\omega_-+\omega_+) + a_-(\omega_--\omega_+)\right],\\
A_+&=& \frac{1}{2}\sqrt{(B_0^2+\lambda)^2-4 K} \left[ a_+ + b_+ \omega_+\omega_+\right], \quad   B_+= \frac{1}{2}\sqrt{(B_0^2+\lambda)^2-4 K} \left[ a_+ - b_+ \omega_-\omega_+\right],\\
k&=& ((B_0^2+\lambda)^2-4 K)^\frac{3}{2} K.
\end{eqnarray*}
We do not write long expressions for $a_\pm$, $b_\pm$.

If we assume that for a characteristic starting from $(x_{01}, x_{02})$
\begin{eqnarray}\label{SC}
C^2&>& A_-^2+B_-^2+A_+^2+B_+^2,
\end{eqnarray}
then the components of $Q$ are bounded. In this way, we obtain a relatively simple sufficient condition for the preservation of smoothness, which does not coincide with the necessary one.

Thus, we obtain the following theorem.

\begin{theorem}\label{T1} 1. The solution of the Cauchy problem \eqref{1}, \eqref{CD1} preserves classical smoothness for all $t>0$ if and only if initial data $\bE_0=\mathcal S_0 \bx$ and $B_0$ are such that for all $(x_{01}, x_{02})\in \mathbb R^2$
${\rm det} Q (t)>0$, where the matrix component $Q=(q_{ij})$ is found as part of the solution to the Cauchy problem
\eqref{ode}, \eqref{CCP} for a linear system with constant coefficients.

2. If for all $(x_{01}, x_{02})\in \mathbb R^2$ the initial data \eqref{CD1}, $\bE_0=\mathcal S_0 \bx$ and $B_0$ are such that
conditions \eqref{mu}, \eqref{nr}, \eqref{SC} are satisfied, then the solution of the Cauchy problem \eqref{1}, \eqref{CD1} preserves classical smoothness for all $t>0$.
\end{theorem}

\begin{remark}
Since in case 2 of Theorem \ref{T1} the function ${\rm det} Q (t)$ is a superposition of two periodic motions with periods $T_1=\frac{2\pi}{\omega_+-\omega_-}$ and $T_2=\frac{2\pi}{\omega_++\omega_-}$,  $T_2<T_1$ (see \eqref{detQ}), then if ${\rm det} Q (t)>0$ for $t\in (0, T_1]$, then ${\rm det} Q (t)>0$ for all $t>0$.
\end{remark}

\subsection{Analysis of the influence of intensity of the magnetic field }\label{S2.1}

Recall that for the case $\bE_0=0$, the necessary and sufficient condition for maintaining the initial smoothness looks very elegant:
\begin{eqnarray*}\label{E0}
(\mathcal D^2 - 4 J +2 B_0 \xi-B_0^2) \Big|_{t=0} <0
\end{eqnarray*}
see  \cite{LT}, \cite{RU}. Thus, if we fix initial data \eqref{CD1} and increase $|B_0|$, we always obtain a globally smooth solution.

For the case $\bE_0\ne 0$, we notice that if we increase $|B_0|$, we obtain a realization of condition \eqref{mu}, so we get the case 2 of  Theorem \ref{T1}.

To trace the influence of $B_0$ in condition \eqref{SC}  and to avoid cumbersome formulae, we consider the axially symmetric case
\eqref{axial}, for which
$s_{11}=s_{22}$, $s_{12}=0$, $v_{11}=v_{22}$, $v_{12}=-v_{21}$. Here the constants in \eqref{SC} look simpler:

\begin{eqnarray*}\label{symconst}
&&C={2F s_{11}^2 B_0^2} (1+ v^2_{11}+v^2_{12}+v_{12} B_0 +B_0^2),
\quad F=B_0\sqrt{B_0^2+4 s_{11}},\\
&&A_-= {s_{11} v_{11}}( B_0^2 + 4 s_{11}) B_0^2( F(\omega_--\omega_+) + B_0^2(\omega_-+\omega_+)), \\
&&B_-= {s_{11} v_{11}} ( B_0^2 + 4 s_{11}) B_0^2( B_0^2(\omega_--\omega_+) -F(\omega_-+\omega_+)), \\
&&A_+= {F B_0^2 s_{11}} (s_{11} (1- v_{11}^2 - v_{12}^2)  - v_{12} B_0) (s_{11}+\omega_-\omega_+) , \\
&&B_+={F B_0^2 s_{11}} (s_{11} (1- v_{11}^2 - v_{12}^2)  - v_{12} B_0) (-s_{11}+\omega_-\omega_+),\\
&&k=( B_0^2 + 4 s_{11})^\frac{3}{2}\,B_0^3 s_{11}^2.
\end{eqnarray*}
It is easy to calculate that for $B_0\to \infty$ $C\sim B_0^6$, while $A_\pm, B_\pm \sim B_0^5$,
so we can get global smoothness that increases $B_0$. We get the same effect for the general case, without the assumption of axial symmetry.

Note also that for $\lambda\to \infty$ (in the axisymmetric case $\lambda=2 s_{11}$) $ C\sim \lambda^\frac{7}{2}$, while $ A_\pm, B_\pm \sim \lambda^\frac{5}{2}$, so another way to get global smoothness is to increase $\lambda$.

%\medskip

\section{Arbitrary $\bE_0$, axially symmetric case}\label{S3}

\medskip

For the axially symmetric solution \eqref{axial} equation \eqref{ODEv} results in
\begin{eqnarray}\label{UV1}
&& \dot U = -U^2+V^2 -B_0 V - S(r), \\\label{UV2}
&&  \dot V = (B_0-2 V) U, \\ \label{UV3}
&&  \dot r = r U.
\end{eqnarray}
Further, since $J={\mathcal D} U +\xi V  U^2 + V^2 + r U U' + r V V'  $, $\mathcal D = 2 U + r U'$, $\xi = 2 V + r V'$, then
$$J={\mathcal D} U +\xi V - U^2 - V^2, $$
 and \eqref{D}, \eqref{xi} can be written as
\begin{eqnarray}\label{Dxi1}
&&  \dot {\mathcal D} =-\mathcal D^2 + 2 \mathcal D U + 2\xi V - 2 U^2 - 2 V^2  - \lambda (r) -B_0 \xi, \\\label{Dxi2}
 && \dot \xi = -\mathcal D (\xi - B_0).
\end{eqnarray}
In this case $\lambda(r)= r S'(r) + 2 S(r)$.
Assume that
\begin{eqnarray}\label{S}
S_- \le S(r)\le S_+,
\end{eqnarray}
where $S_\pm$ are constants.

\bigskip

\subsection{Behavior of the solution}\label{S3.1}

1. If $S(r)=S_0 =\rm const$, i.e. in the case of an affine $\bE_0$ considered in the previous section,
the system consisting of \eqref{UV1}, \eqref{UV2} can be explicitly integrated. Namely, the phase curve on the plane $(U,V)$ is a circle,
\begin{eqnarray}\label{phase}
&&U^2 +(V+(C_1-\frac{B_0^2}{4}))^2 = (C_1+\frac{B_0^2}{4}))^2 - S_0 - \frac{B_0^2}{4},\\
&&C_1=\frac{1}{4} \left(\frac{4S_0+B_0^2 + 4 U_0^2+ 4 V_0^2 -2 B_0 V_0}{B_0-2 V_0}  \right),\quad U_0=U(0), \, V_0=V(0),\, V_0\ne \frac{B_0}{2}.\label{C1}\nonumber
\end{eqnarray}
System \eqref{UV1}, \eqref{UV2} for the case $S(r)=S_0 =\rm const$ has the following equilibria:
\begin{itemize}
\item $U=0, \quad V=\frac12 B_0\pm \frac12\sqrt{4 S_0+ B_0^2}, $ for $4 S_0+ B_0^2= 2 \lambda+ B_0^2>0$,
centers, period of revolution along every phase curve is $T=\frac{2 \pi }{\sqrt{4 S_0+ B_0^2}}$;

\item $U=\frac12\sqrt{4 S_0+ B_0^2}, \quad V=\frac12 B_0, $ for $4 S_0+ B_0^2= 2 \lambda+ B_0^2<0$,
stable and unstable nodes (which degenerate at $4 S_0+ B_0^2=0$).
\end{itemize}

\bigskip

2. For arbitrary smooth $S(r)$ equations \eqref{UV2}, \eqref{UV3} imply
$$r= \frac{C_2}{\sqrt{|-2 V + B_0|}}, \quad C_2=r_0 \sqrt{|-2 V_0 + B_0|}, $$
therefore $S(r)=S(V)$ and the phase curve of \eqref{UV1}, \eqref{UV2} takes the form
\begin{eqnarray}\label{phase1}
&&U^2 +(B_0-2V)\left(-\frac12 V+C_3\right) + G(V)= \frac{B_0^2}{4},\\
&&G(V)=2 (B_0-2V)\int\limits_{\infty}^V \frac{S(\nu)}{(B_0-2\nu)^2} \,d \nu.\nonumber
\end{eqnarray}
Since for $S(r)\in [S_-, S_+]$
$$
S_-\le G(V) \le S_+,
$$
then the phase curve \eqref{UV1}, \eqref{UV2} lies between the two \eqref{phase} circles corresponding to $S_-$ and $S_+$, where the constants $C_1$ and $C_3$ calculated with the same initial data $(U_0, V_0)$.

\begin{remark}
The integral $G(V)$ can be found explicitly for many important choices of $S(r)$, for example, $\sin r$, $\cos r$, $\frac{1}{1+r^\alpha}$, $\alpha= 1, 2, 3, 4,$ etc.
\end{remark}

\bigskip

Since we want to obtain an analogue of Theorem  \ref{T1}, we will focus on the first case $ 2 \lambda+ B_0^2 > 4 S_-+ B_0^2>0$ (this condition corresponds to \eqref{mu}) .

%Note that in the axisymmetric case $\lambda= r S'(r)+2 S(r)$.

\begin{lemma}\label{L1} Let condition \eqref{S} holds and $4 S_-+ B_0^2>0$.
Then the solution $(U,V,r)$ of the Cauchy problem
\eqref{UV1}, \eqref{UV2}, \eqref{UV3},
\begin{eqnarray*}
(U, V, r) \Big|_{t=0} = (U_0, V_0, r_0),
\end{eqnarray*}
is bounded from above and below by constants that depend on the initial data. Namely,
\begin{eqnarray}\label{bound}
U_-\le U \le U_+, \quad V_-\le V \le V_+,
\end{eqnarray}
where
$$
U_\pm= \pm \max\{R_-, R_+\}, \qquad V_\pm=\frac14 B_0 \pm \max\{ -c_-\pm R_-, -c_+\pm R_+\},
$$
$$
c_\pm=\frac{1}{4} \left(\frac{4S_\pm+B_0^2 + 4 U_0^2+ 4 V_0^2 -2 B_0 V_0}{B_0-2 V_0}  \right),
$$
and
$$R_\pm^2=(c_\pm+\frac{B_0^2}{4}))^2 - S_\pm - \frac{B_0^2}{4},\quad R_\pm>0.$$
%If $V_0>\frac12 B_0+ \frac12\sqrt{4 S_0+ B_0^2}=\mathcal V_+$ or $V_0<\frac12 B_0- \frac12\sqrt{4 S_0+ B_0^2}=\mathcal V_-$, $U_0= 0$. then
\end{lemma}

\proof
First of all, let us note that \eqref{phase1} implies that the phase curve of system \eqref{UV1}, \eqref{UV2} is symmetric with respect to the axis  $U=0$ and the axis $V=\frac12 B_0$, (the equations do not change for $U_1=-U$ and $V_1=B_0-V$) therefore we can consider only the quadrant $U\ge 0$, $V>\frac12 B_0$.

From  \eqref{UV1}, \eqref{UV2} we have
\begin{eqnarray*}%{l}
\frac{d U}{d V}=\frac{-U^2+V^2 -B_0 V - S(r)}{-U (2 V-B_0)}=\Psi(Z, V, t),
\end{eqnarray*}
or
\begin{eqnarray}\label{ZV}
\frac{d Z }{d V}=\frac{-Z^2+V^2 -B_0 V - S(r)}{-(2 V-B_0)},\quad Z=
\frac12 U^2.
\end{eqnarray}
Let us denote
$$\Psi_\pm(Z, V, t)= \frac{-Z^2+V^2 -B_0 V - S_\pm}{-(2 V-B_0)}.$$

Since $V>\frac12 B_0$,
$$\Psi_-(Z, V, t)\le \Psi(Z, V, t)\le \Psi_+(Z, V, t).$$
Now we can apply Chaplygin's theorem on differential inequalities,
according to which the solution $ Z (V) $ of the Cauchy problem for
\eqref{ZV} with initial conditions $ Z (V_0) = Z_0 $ for $ V >V_0 $
satisfies the inequality
\begin{eqnarray*}\label{Z_ineq1}
  Z_-(V) \le Z(V,t)\le  Z_+(V),
\end{eqnarray*}
and for $V<V_0$ the inverse inequality
\begin{eqnarray*}\label{Z_ineq2}
  Z_-(V) \le Z(V, t)\le  Z_+(V),
\end{eqnarray*}
where $Z_\pm (s)$ are the solutions to problems $ \frac{d Z}{d V}=\Psi_\pm (Z,V)$, $Z(V_0)=Z_0 $.

Thus, for  $V<V_0$ we have
 $Z(V, t)\ge  Z_-(V)$, for  $V>V_0$ we have
 $Z(V, t)\ge  Z_+(V)$, $U=\sqrt {2 Z} \ge 0$.  The period $T$ of motion along the phase curve can be estimated as $\frac{2 \pi}{\sqrt{4 S_++B_0^2}}\le T\le
\frac{2 \pi}{\sqrt{4 S_-+B_0^2}}$.

The behavior of the phase curves is shown in Fig.1.
$\Box$

%\begin{center}
\begin{figure}[htb]
\begin{center}
\includegraphics[scale=0.4]{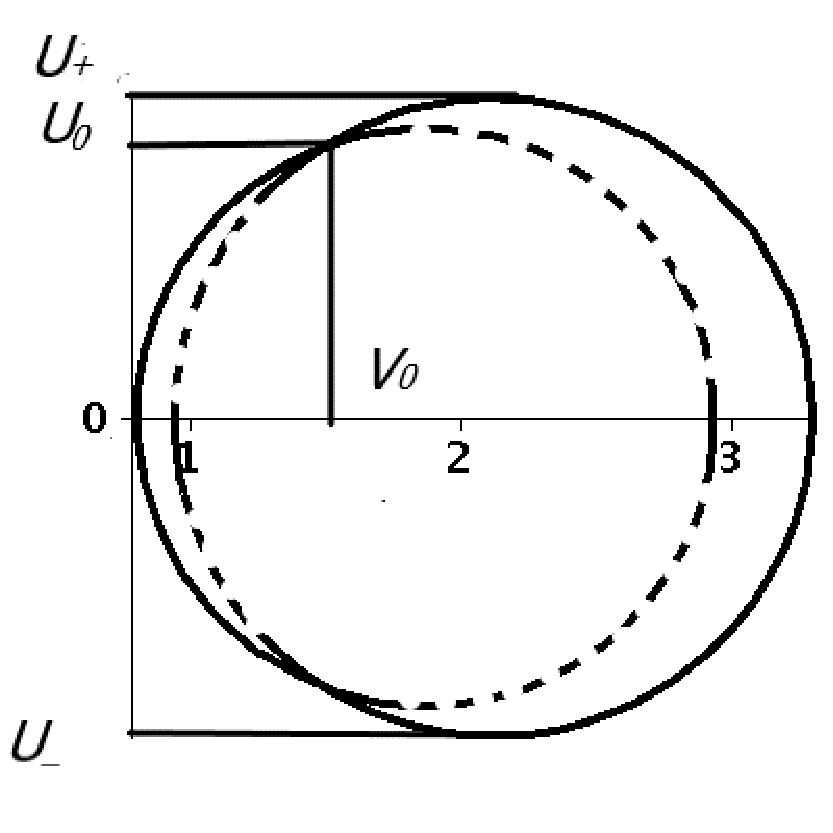}
%\vspace{-0.5 cm}
\end{center}
\caption{Graphs for $U=\pm \sqrt{2 Z_\pm}$. Combination of graphs limiting the phase curve, solid line,  for $1<S(r)<2$,  $B_0=1,$ $U_0=1$, $V_0=1.5$.}
\end{figure}
%\end{center}

\medskip

\subsection{Behavior of the derivatives}\label{S3.2}

Now we can study the behavior of the divergence and vorticity of the solution. Recall that, by the properties of hyperbolic systems, the boundedness of $\mathcal D$ and $\xi$ implies that the solution of the Cauchy problem \eqref{1}, \eqref{CD1} preserves the original smoothness \cite{Daf}.

If we change $\eta=\xi-B_0$,  system \eqref{Dxi1}, \eqref{Dxi2} can be rewritten as
\begin{eqnarray}\label{Deta1}
 && \dot {\mathcal D} =Y(\mathcal D, \eta, U, V, \lambda)= -\mathcal D^2 + 2 \mathcal D U + \eta (2 V-B_0)  - 2 U^2 - 2 V^2 -2 B_0 V  -B^2_0- \lambda, \\
 && \dot \eta = -\mathcal D \eta.\label{Deta2}
\end{eqnarray}
As follows from the results of Sec.\ref{S3.1}, $\lambda(r)=\lambda(V)$ is a periodic function. Let us assume
\begin{eqnarray}\label{lambda}
\lambda_- \le \lambda(r)\le \lambda_+,
\end{eqnarray}
where $\lambda_\pm$ are constants.

\medskip

1. System \eqref{Deta1}, \eqref{Deta2} can be linearized by means of the Radon lemma (Theorem \ref{T2}). Indeed,
here
%$$
\begin{eqnarray*}\label{M}
W=\begin{pmatrix}
  \mathcal D\\
  \eta
\end{pmatrix},\quad
M_{11}=\begin{pmatrix}
  0\\
\end{pmatrix},\quad
 M_{12}=\begin{pmatrix}
  1 & 0\\
\end{pmatrix},\\
M_{21}=\begin{pmatrix}
 G \\ 0
\end{pmatrix}, \quad
M_{22}=\begin{pmatrix}
  2UF& 2V-B_0\\
  0 & 0\\
\end{pmatrix},\\\nonumber
 G=- 2 U^2 - 2 V^2 -2 B_0 V  -B^2_0- \lambda.
\end{eqnarray*}

Thus, we obtain the linear Cauchy problem
\begin{eqnarray}
\label{matr}
 \begin{pmatrix}
  \dot q\\
  \dot p_1\\
  \dot p_2\\
\end{pmatrix}
=\begin{pmatrix}
0& 1& 0\\
G& 2\,U& 2\,V-B_0\\
0& 0  & 0\\
\end{pmatrix}
\begin{pmatrix}
  q\\
  p_1\\
  p_2\\
\end{pmatrix},\quad
\begin{pmatrix}
  q\\
  p_1\\
  p_2
  \\
\end{pmatrix}(0)=\begin{pmatrix}
  1\\
  \mathcal D_0\\
  \eta_0\\
\end{pmatrix},
\end{eqnarray}
with periodical coefficients, known from \eqref{UV1} -- \eqref{UV3}. System
\eqref{matr} implies %the linear ODE with respect to to $p_1$:
\begin{eqnarray}\label{q}
  \ddot q-2 U \dot q -G q=(2V-B_0)\eta_0, \quad q(0)=1,\, \dot q(0)=\mathcal D_0,
\end{eqnarray}
which can be written as
\begin{eqnarray}\label{y}
  &&\ddot y + (3V^2+V B_0 + 2 B_0^2 -S(V) + 2 \lambda(V)) y=(2V-B_0)\eta_0 e^{-\int\limits_0^t U(\tau) d\tau}, \\&& y(0)=1,\, \dot y(0)=\mathcal D_0-U_0,\quad y=q e^{-\int\limits_0^t U(\tau) d\tau},\nonumber
\end{eqnarray}
and the solution to \eqref{Deta1}, \eqref{Deta2} blows up if and only if the solution to \eqref{q} (and \eqref{y}) vanishes.

As follows from the results of Sec.\ref{S2}, for $S=S_0=\rm const$ if a blow up happens, it happens on the first period of oscillation,
however, in the case of a general form of $S(r)$ the solution of \eqref{y} can be resonant and the amplitude of oscillations can increase.
%Nevertheless, we are going to show that one can find conditions on the initial data, guaranteeing the global smoothness of the solution.

\medskip

2. Let us find a sufficient condition for the preservation of smoothness in the first period of oscillations $T\le
\frac{2 \pi}{\sqrt{4 S_++B_0^2}}$.

We assume $V_0>\frac{B_0}{2}$, $\eta>0$ and obtain two-sided estimates.
\begin{eqnarray*}\label{Y}
  Y\le Y_{1+}=-\frac34 \mathcal D^2 + \eta^2 + K_{11}, \qquad K_{11}=    2 (U_+^2 + V_+^2 - B_0 V_-)  - \lambda_-, \\
  Y\le Y_{2+}=-\frac34 \mathcal D^2 + a_+ \eta + K_{12}, \qquad K_{12}=   3 U_+^2 - 2  V_+^2 -2 B_0 V_+  -B^2_0- \lambda_+,\quad a=2V_--B_0,\\
  Y\ge Y_-=-\frac54 \mathcal D^2 + a_- \eta + K_2, \qquad K_2=   -6 U_+^2 - 2  V_+^2 -2 B_0 V_-  -B^2_0- \lambda_+,\quad a=2V_--B_0,
\end{eqnarray*}
Thus, with the change $ \mathcal Z =\frac12 \mathcal D^2 $
\begin{eqnarray}\label{Zeta}
\frac{d \mathcal Z}{d \eta}=\frac{Y}{-\eta}=\Phi(\mathcal Z, \eta, U, V, \lambda).
\end{eqnarray}

Similar to Sec.\ref{S3.1} we denote
$$\Phi_\pm(\mathcal Z, \eta)= \frac{Y_\mp}{-\eta},$$
therefore
$$\Phi_-(\mathcal Z, \eta)\le \Phi(\mathcal Z, \eta, t)\le \Phi_+(\mathcal Z, \eta).$$
Thus, the Chaplygin's theorem implies that
according to which the solution $ \mathcal Z (V) $ of the Cauchy problem for
\eqref{Zeta} with initial conditions $ \mathcal Z (\eta_0) = \mathcal Z_0 $ for $ \eta >\eta_0 $
satisfies the inequality
\begin{eqnarray*}\label{Z1_ineq1}
  \mathcal Z_-(\eta) \le \mathcal Z(\eta,t)\le \mathcal Z_+(\eta),
\end{eqnarray*}
and for $\eta <\eta_0$ the inverse inequality
\begin{eqnarray*}\label{Z1_ineq2}
 \mathcal Z_+(\eta) \le \mathcal Z(\eta,t)\le \mathcal Z_-(\eta),
\end{eqnarray*}
where $\mathcal Z_\pm (\eta)$ are the solutions to problems $ \frac{d \mathcal Z_\pm}{d \eta}=\Phi_\pm (\mathcal Z,\eta)$, $\mathcal Z_\pm(\eta_0)=\mathcal Z_0 $.

For $\eta_0>0$, $\mathcal D_0=\sqrt{2 \mathcal Z_0}\ge 0$, $\mathcal Z$ decreases, %(see Fig.2),
therefore $\eta<\eta_0$ and
 $\mathcal Z_+(V)\le \mathcal Z(\eta, t)\le \mathcal Z_-(V)$,  up to the point $0<\eta_{00}\le\eta_+$, where $\eta_+$ is the smaller of the solutions of $ \mathcal Z_+(\eta)=0$.
 Then we take the point $(\eta_{00}, 0)$ as a new initial data, in the semi-plane $\mathcal D<0$ the value of $\eta$ increases
 and $\mathcal Z(\eta, t)\le  \mathcal Z_+(\eta)$, $\mathcal D_0=-\sqrt {2 \mathcal Z}_0 \le 0$.% Thus, for the case $\mathcal D_0=0$ we have to make only one switch from $\mathcal Z_-$ to $\mathcal Z_+$.

 It is easy to see that the curve,  $\mathcal D_+ =\mathcal D_+(\eta)$,
 which bounds the phase curve to \eqref{Deta1}, \eqref{Deta2} from above for $\mathcal D>0$ (with the estimate by means of $Y_{1+}$) is given by
 $$\mathcal D_+^2+4\eta^2 - C_+ \eta^\frac{3}{2}=\frac43 K_1,$$
 where the constant $C_+$ is defined by the initial point $(\mathcal D_0>0, \eta_0>0)$ and it is bounded for any $C_+$ (the oldest degree of $\eta$ is 2.) This means that the divergence $\mathcal D$ cannot blow up in the upper half-plane. From the other side, the curve $\mathcal D_- =\mathcal D_-(\eta)$, which bounds the phase curve to \eqref{Deta1}, \eqref{Deta2} from below for $\mathcal D<0$, is given by
 $$\mathcal D_-^2-\frac43 a_-\eta - C_- \eta^\frac{5}{2}=\frac45 K_2, $$
 where the constant $C_-$ is defined by the initial point $(\mathcal D_0\le 0, \eta_0>0)$ and it is bounded only if  $C_-<0$ (the oldest degree of $\eta$ is $\frac52$.) Thus, the initial data corresponding to the condition $C_-<0$ is
 \begin{eqnarray}\label{cond_bl}
 \mathcal D_0^2-\frac43 (2V_--B_0) \eta_0 < -6 U_+^2 - 2  V_+^2 - 2 B_0 V_+  -B^2_0- \lambda_+ ,\quad \mathcal D_0< 0,
  \end{eqnarray}
 the values of $U_+, \, V_\pm$ are written  in \eqref {bound}, Lemma \ref{L1}.

 The case $\xi<0$ can be considered analogously.

  %If $\mathcal D_0>0$, then $\mathcal D_-^2-\frac45 a \eta - C_- \eta^\frac{5}{2}=\frac45 K_2 $ is a lower bound for the phase curve and therefore the trajectory comes to the point with the coordinate $\mathcal D_1=-\mathcal D_0$ and $0<\eta_1<\eta_0$. Therefore, condition  \eqref{cond_bl} guarantees the boundedness of $\mathcal D$ during the first rotation of the trajectory for all signs of $\mathcal D_0$.

  \medskip

 The following theorem sums up our reasoning.

  \begin{theorem}\label{T3} Consider the Cauchy problem \eqref{1}, \eqref{CD1} for the axially symmetric class of solutions \eqref{axial}
  and assume that
  the fixed field $\bE_0$ is such that conditions \eqref{S} and \eqref{lambda} hold
  for all $r_0\in \overline{\mathbb R}_+$, and
  $U_0(r),\, V_0(r),\, \Div \bV_0=\mathcal D_0,\, \Rot \bV_0=  \xi(r) $ are such that condition \eqref{cond_bl} is valid for all $r_0\in \overline{\mathbb R}_+$.
  Then the time $T$ of existence of the classical solution to the Cauchy problem can be estimated from below as
  \begin{eqnarray}\label{Time}
   T\le
\frac{2 \pi}{\sqrt{4 S_++B_0^2}}.
  \end{eqnarray}
 \end{theorem}

Fig.2 shows estimates of phase trajectories in the upper and lower half-planes for $\mathcal D$.

\begin{center}
\begin{figure}[htb]
%\hspace{1cm}
\begin{minipage}{0.4\columnwidth}
%\centerline{
\includegraphics[scale=0.3]{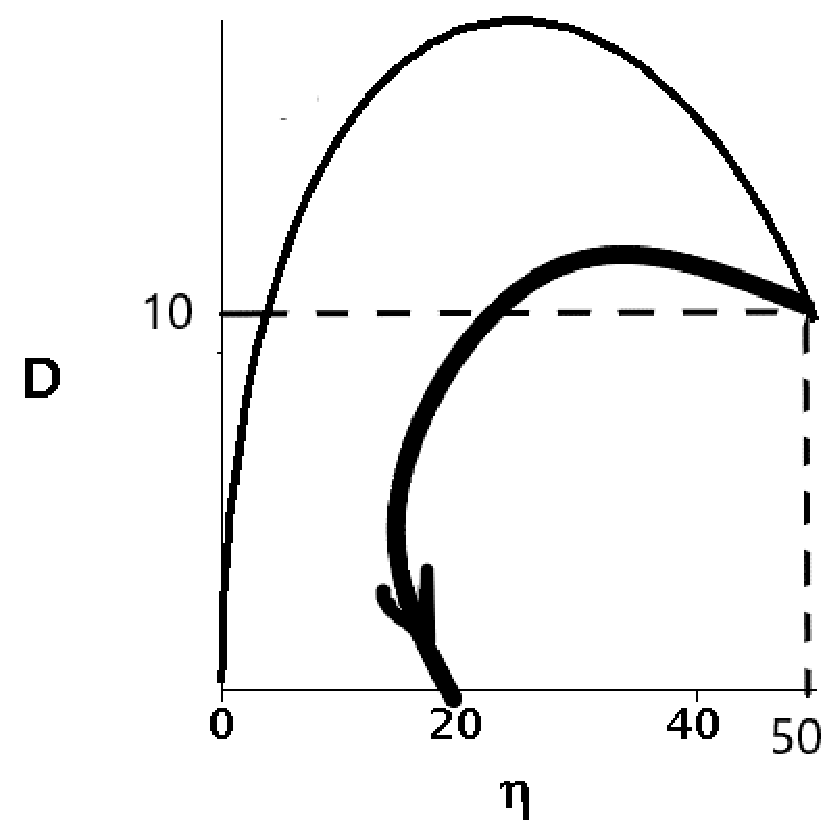}
%\vspace{-0.5 cm}
\end{minipage}
\hspace{1cm}
\begin{minipage}{0.4\columnwidth} %\centerline{
\includegraphics[scale=0.3]{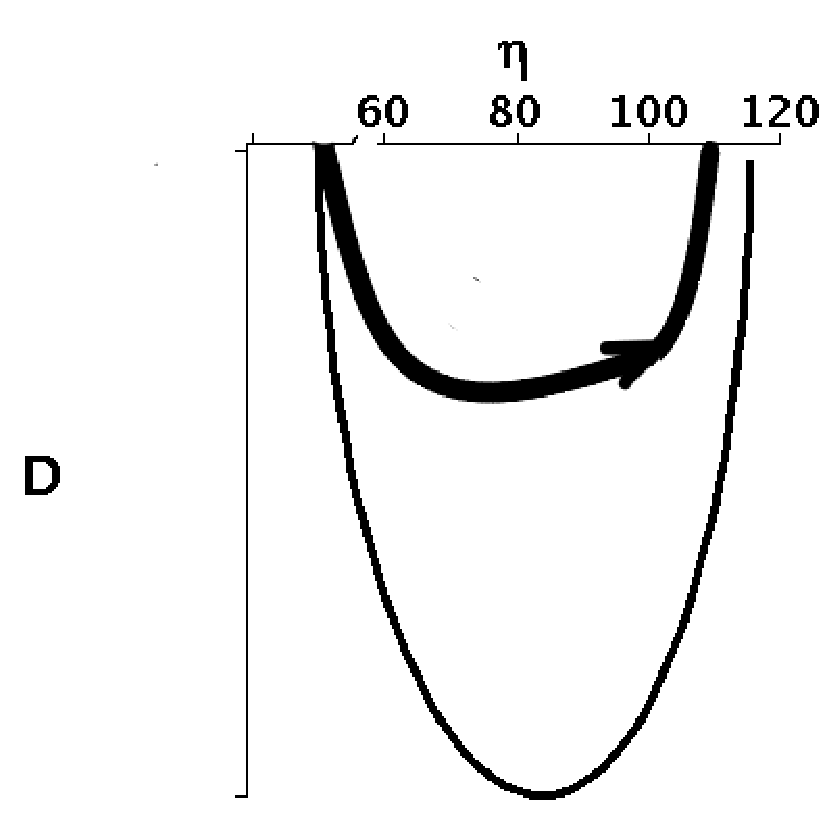}
%\vspace{-0.5 cm}
\end{minipage}
\caption{Left: graphs for $\mathcal D= \sqrt{2 \mathcal Z_-}(\eta)$, upper bound for the phase trajectory, $\mathcal D>0$, the initial point is $\mathcal D_0=10,$ $\eta_0=50$. Right: graphs for $\mathcal D=- \sqrt{2 \mathcal Z_+}(\eta)$, lower bound for the phase trajectory, $\mathcal D<0$, the initial point is $\mathcal D_0=0,$ $\eta_0=50$; the trajectory returns to the upper half-plane. Here $U_+ = 1,  V_+ = 5, V_- = 1, B_0= 1, \lambda_- = -1, \lambda_+ = 1$.
  }\label{FPic}
\end{figure}
\end{center}

  \begin{remark} In the proof of Theorem \ref{T3}, rough and simple estimates $Y(\mathcal D, \eta, U, V, \lambda)$ are used, so the sufficient condition for maintaining smoothness is far from being exact. The absence of a bounded curve $\mathcal Z_+$ for specific initial data in the lower half-plane $\mathcal D<0$ does not mean that the phase trajectory goes to infinity.
    The lower estimate \eqref{Time} is also very rough, and we can continue counting the number of revolutions by following the algorithm \cite{Roz22_MMAS}.
    \end{remark}

  \begin{remark} Note that a large initial vorticity helps the implementation of \eqref{cond_bl} with all other parameters fixed.
 \end{remark}

 \begin{remark} A very interesting problem, which, it seems, can only be solved numerically, is the calculation of the Floquet multipliers for the linear system \eqref{matr}, see \cite{RTur}, for various  landscapes of $\bE_0$. This would help answer the question whether we can control the smoothness of the solution and the stability of the equilibria using $\bE_0$.
  \end{remark}

%\section*{Acknowledgements} Supported by RSF grant 23-11-00056 through RUDN University.


\begin{thebibliography}{99}
\bibitem{ABR78} A. F. Alexandrov, L. S. Bogdankevich, A. A. Rukhadze, {\it  Principles of plasma electrodynamics,  Springer
series in electronics and photonics}, Springer: Berlin Heidelberg
(1984).


\bibitem{CH18}
E. V. Chizhonkov, {\it Mathematical aspects of modelling oscillations
and wake waves in plasma}, CRC Press (2019).


\bibitem{GFCA} L. M. Gorbunov, A. A. Frolov, E. V. Chizhonkov,
N. E. Andreev, {``Breaking of nonlinear cylindrical plasma
oscillations''}, {\it Plasma Physics Reports, \bf 36} (4), 345--356 (2010).




\bibitem{Daf}
 C. M. Dafermos, {\it Hyperbolic conservation laws in continuum physics}, The 4th Edition, Berlin-Heidelberg: Springer (2016). %852 P.


\bibitem{david72}
R. C. Davidson, {\it Methods in Nonlinear Plasma Theory}, Acad. Press, New
York (1972).







\bibitem{esarey09}E. Esarey, C. B. Schroeder,  W. P. Leemans, ``Physics of laser-driven plasma-based
electron accelerators'', {\it Rev. Mod. Phys.,  \bf 81},
1229--1285 (2009).

\bibitem{Riccati} G. Freiling,
``A survey of nonsymmetric Riccati equations'', {\it Linear Algebra and its
Applications, \bf 351-352}, 243--270 (2002).

\bibitem{GR75} V. L. Ginzburg, {\it Propagation of electromagnetic waves in plasma}, Pergamon, New York (1970).

\bibitem{LT} H .Liu, E. Tadmor, ``Rotation prevents finite-time breakdown'', {\it Physica D: Nonlinear Phenomena, \bf 188} 262--276 (2004).


\bibitem{Radon} {W. T. Reid,}
{\it Riccati differential equations}, Academic Press, New York  (1972).

\bibitem{RChZAMP21}
O. S. Rozanova, E. V. Chizhonkov, ``On the conditions for the
breaking of oscillations in a cold plasma'', {\it Z. Angew. Math. Phys.,
 \bf 72}, 13 (2021). %DOI: /10.1007/s00033-020-01440-3.

\bibitem{RChZAMP22} O. S. Rozanova, E. V. Chizhonkov,
``The influence of an external magnetic field on cold plasma oscillations'', {\it Z. Angew. Math. Phys., \bf 73}, 249 (2022).
%doi:/10.1007/s00033-021-01615-6.



\bibitem{R_PhysicaD22} O. S. Rozanova, ``On the behavior of multidimensional radially symmetric solutions of the repulsive Euler-Poisson equations'', {\it Physica D: Nonlinear Phenomena, \bf 443}, 133578 (2022).

 \bibitem{Roz22_MMAS} O. S. Rozanova,   ``On the properties of multidimensional electrostatic oscillations of an electron plasma'', {\it Math. Meth. Appl. Sci.
  \bf 46}, 7557--7571 (2023).

  \bibitem{RU} O.~S.~Rozanova, O. V. Uspenskaya, ``On properties of solutions of the cauchy problem for two-dimensional transport equations on a rotating plane'',
{\it  Moscow University Mathematics Bulletin, \bf 76} (1),  1--8 (2021).

 \bibitem{RTur} O.~Rozanova, M.~Turzynsky, ``On the properties of affine solutions of cold plasma equations'',
{\it Communications in Mathematical Sciences, \bf 21} (2023), in press.





\bibitem{Shep13} { C. J. R. Sheppard,}
``Cylindrical lenses --- focusing and imaging: a review'', {\it Applied Optics,  \bf 52}, 538--545 (2013).



\end{thebibliography}
\end{document}